\documentclass[11pt, a4paper]{article}
\usepackage{amsmath}
\usepackage{amsfonts}
\usepackage{amssymb}
\usepackage{dirtytalk}
\usepackage{graphicx}
\usepackage{hyperref}
\usepackage[top=3.17cm, bottom=2.54cm, left=2.54cm, right=2.54cm, includehead, includefoot,]{geometry}
\usepackage{authblk}
\usepackage{textcomp}
\usepackage{natbib}
\usepackage{float}
\usepackage{graphicx}

\begin{document}
\title{\vspace{-3.0cm}\large\textbf{Decentralising the United Kingdom: the Northern Powerhouse strategy and urban ownership links between firms since 2010}}
\author[1,2,3*]{\footnotesize Natalia Zdanowska}
\author[2]{\footnotesize Robin Morphet}
\affil[1]{\footnotesize Luxembourg Institute of Socio-Economic Research, Luxembourg}
\affil[2]{\footnotesize Centre for Advanced Spatial Analysis, University College London, London, United Kingdom}
\affil[3]{UMR CNRS 8504 G{\'e}ographie-cit{\'e}s, Paris, France}
\affil[*]{n.zdanowska@ucl.ac.uk}
\date{\vspace{-5ex}}

\maketitle
\thispagestyle{empty}

\noindent
    
    This paper explores a decentralisation initiative in the United Kingdom -- the Northern Powerhouse strategy (NPS) -- in terms of its main goal: strengthening connectivity between Northern cities of England. It focuses on economic interactions of these cities, defined by ownership linkages between firms, since the NPS\textquotesingle s launch in 2010. The analysis reveals a relatively weak increase in the intensity of economic regional patterns in the North, in spite of a shift away from NPS cities\textquotesingle  traditional manufacturing base. These results suggest potential directions for policy-makers in terms of the future implementation of the NPS. \newline
    

\section*{Introduction}

    According to Sharon White, then (20th January 2015) second permanent secretary at HM Treasury, the United Kingdom is \say{almost the most centralised developed country in the whole world} \citep{agbonlahor_2015} thus identifying one of the most important issues facing the UK. This  reflected the 2014 view of OECD chief executive Alexandra Jones who said that within the OECD, the UK \say{is one of the most centralised countries} \citep{bbcJones2014,charbit2011governance}.  Strong centralisation has not always characterised the British urban system. In the 19th century  Manchester together with Liverpool was one of the most industrialised conurbations in Europe, a major textile producer and, in 1830, the first to have passenger railway connections \citep{ElbaumLazonick1984}.  Manchester, the UK\textquotesingle s second economic centre, has a significant history in terms of innovation and social progress and, together with Liverpool, was the first identified conurbation \citep{geddes1915cities}. \newline

    Despite many decentralisation initiatives proposed by the UK government to devolve power to city regions \citep{Sandford2020}, sub-regional governance remains subordinate to a centralised model of decision making. The partial move of the BBC from London to Manchester is one example of a decentralising governmental initiative. Other examples such as  City Deals \citep{O'BrienPike2015}  the establishment of Combined Authorities and the introduction of City Mayors \citep{Sandford2020} must be seen in the context of the loss of spending power of English Metropolitan Districts of over 20\%  between 2009/2010 and 2018/19 \citep{atkins2020} as a result of reductions in the Revenue Support Grant. Decentralisation is a central challenge particularly in the context of the revenge of \say{the places that don’t matter} \citep{RodriguezPose2017}, of rising populism and inequalities in the country \citep{Carrascal--Inceraetal2020}. The Northern Powerhouse (NP) together with the Northern Powerhouse strategy (NPS) is a central government proposal, albeit one with strong local support.  It was launched in 2010, to boost economic growth in the North of England particularly in the core cities of Manchester, Liverpool, Leeds, Sheffield, Hull and Newcastle. It corresponds to a vision of a globally connected and competitive Northern Economy, exploiting strengthened connections between and within the Northern cities \citep{NorthernPowerhouse2016}. The Prime Minister \citep{johnson2019speech}  pledged support for the NP saying  \say{We are going to maximise the power of the North with more mayors across the whole of the North}. The NP Partnership Plan for 2050 claims that it will improve the productivity of the North in its vision for the future \citep{NorthernPowerhouse2016}. Decentralising Britain is considered as the \say{big push} towards inclusive prosperity \cite{Esteveetal2019} and reflects the view that, to counterbalance the forces of growing globalisation, local economic strengths should emerge as a countervailing force \citep{Roudometof2017}.
    
    This paper aims at exploring these local strengths \citep{Roudometof2017} within the NP economy\footnote{Cheshire and Warrington, Cumbria, Greater Manchester, Humber, Lancashire, Leeds City Region, Liverpool City Region, North East, Sheffield City Region, Tees Valley, York, North Yorkshire and East Riding. North Wales is defined as covering: Wrexham, Rhyl, Colwyn Bay, Llandudno and Bangor.}. The original contribution of this paper is to approach these issues within the system of cities framework \citep{Pumain1997}. In the latter we consider only the theoretical aspects where the position and dynamics of cities in their spatial and socio-economic dimensions are considered in the context of their interactions with other cities. Preferential interactions between certain cities of a national system can drive  the emergence of sub-systems, which can be relevant to test in the case of the NP economy. In addition, strengthening the national system of cities and building relationships and partnerships between cities are all outcomes of foresighting defined by the UK government for the next fifty years \citep{LandUseFutures}. In fact recent studies have stressed the importance of urban foresight \citep{Dixonetal2018a,DixonTewdwr-Jones2021} and urban foresight techniques for city visioning \citep{Dixonetal2018b} in order to create opportunities for new investment in the local urban economy and to elaborate strategic urban intelligence \citep{RavetzMiles2016}. From this perspective, a particular focus will be given to the city of Manchester, to evaluate its regional role as the second ranked economic centre of the UK. To conduct this research, we investigate transnational firm networks defined by ownership links between firms at city-level, using the Bureau Van Dijk\textquotesingle s ORBIS and AMADEUS databases. We question how Manchester\textquotesingle s position has evolved within its inter-urban economic networks in a highly centralized city-system, following the launch of the NP initiative in 2010 \citep{Coxetal2016}.
    
    The paper is divided into four sections: literature review, hypotheses, materials and methods, results and discussion.

\section*{Literature review}

    Local embeddedness in economic geography has tended to conflate network and territorial dimensions in focusing on the nature of a firm\textquotesingle s linkages to, and interactions with, other local actors, whether firms or cities. These interactions are crucial for regional development and can contribute to revitalizing their surrounding regions \citep{Batty2013,Clarketal2018}. Strengthening these inter-urban ties is one of the NP\textquotesingle s objectives, although no clear critical assessment of this policy in terms of economic interactions has been made since 2010. This obscures a clear city-vision for the NPS as a whole, as  has already been  constructed for separate UK cities such as Reading \citep{Dixonetal2018b} or Newcastle \citep{TewdwrJonesetal2015} with the aim of identifying those factors which could underpin a strong local economy in the future. In this context urban (or city) foresight is understood as the science of thinking about the future of cities. "It draws on diverse methods to give decision-makers comprehensive evidence about anticipated and possible future change" \citep{Dixonetal2018b}. In this sense urban foresight in the present paper is conducted by analysing the NP\textquotesingle s  economic interactions between firms.
    
    There are many reasons why a firm might want to possess a subsidiary in a location remote from the parent firm. The parent firm may be seeking to ring fence a particular operation \citep{bateswells2019} isolating both risks and assets. For example, a subsidiary may be a locally well known brand or the subsidiary may be developing innovative products with their concomitant risks. The extent of ring fencing may vary according to whether or not the parent company has guaranteed or indemnified its subsidiary. It may also vary in terms of subsidiary autonomy from the parent. Autonomy may range from hands on local management control to shared corporate policies or more remote performance management. The latter may compare performance with other subsidiaries of the parent rather than with more local competition. The subsidiary sits somewhere in the competitive spectrum between the parent institutional economy \citep{Coase37} and the more or less, free market of its local economy, and may seek to benefit from both. Investment in subsidiaries by foreign firms may also take into account taxation benefits and the relative stability of trading conditions under a different jurisdiction. 
    
    The rationale \citep{marshall30,ellison2010causes} for agglomeration economies concentrates on the role of supply factors such as skilled labour, knowledge spillovers and transport costs and the ability to access scale economies. These are for the most part, treated statically. The advantages, however, of a subsidiary organisation which accrue to a parent firm may include its access to more dynamic factors such as a continually improving pool of skilled labour, access to knowledgeably up to date capital investment, access to an increasingly specialised supply chain of non-traded intermediates and to an informed market \citep{porter1994role}. The positive feedback of the  forward and backward linkages \citep{krugman91a} which may be exploited by a relatively autonomous subsidiary may well be sufficient to outweigh the advantages of vertical linkages stemming from locating the subsidiary activity within the parent firm. Another strong reason for locating in such an environment is the need to develop trusted bilateral relationships \citep{mignot2020market}  when progressing the development of innovative products. Bilateral relationships in such markets require a degree of trust that is unnecessary in auction markets where price is the dominant component of information exchange.  \cite{granovetter2005impact} describes the characteristics of such networks in terms  of norms and network density. Dense networks of contacts tend to reinforce norms of conduct and the application of sanctions for nonconformity as against the use of impersonal auction markets for supply. \cite{gammelgaard2012impact} found, when examining subsidiaries of foreign firms, that increased autonomy improved inter organisational networks and that increased inter organisational networks improved subsidiary performance. A less sanguine view of the benefits of autonomy is given by  \cite{puck2016ownership}  who compared foreign owned subsidiaries  with international joint ventures in China and found that the wholly owned foreign subsidiaries  performed better. They argue the need to differentiate control from ownership since control can vary considerably within similar levels of ownership. They go on to argue that cultural distance will also affect the level of control reducing it for wholly owned subsidiaries and increasing it for joint ventures. Of course the cultural distance between China and a foreign firm will be rather larger than the cultural distance between London and the NP but the latter should not therefore be ignored. 
 
    Subsidiary performance is an active field of research which has relied largely on cross sectional surveys. Autonomy and agglomeration benefits are recognised as important components of cluster formation. Less studied is  the temporal dimension although its  importance is acknowledged \citep{avnimelech2010regional}. Considering these structures and relationships  begs the question of what sort of evolution of  arrangements we might expect to see in a NP which was successfully transitioning into  a counterweight to London. We might follow the analysis of \cite{avnimelech2010regional} for subsidiaries set up by foreign firms. In this case we would argue for the support  of  inward investment through subsidiary firm location on the grounds that this would encourage the development of further more locally owned subsidiaries or start-ups as off shoots of the foreign owned subsidiaries. The new firms would benefit from the entrepreneurial and management experience gained in the externally owned subsidiaries.  Conversely we might follow the analysis of manufacturing \citep{ellison2010causes} which emphasised the Marshallian agglomeration factors of the movement of ideas, goods and people and the need to reduce the cost of such movements. In practice both will be necessary. Transport improvements are not enough by themselves but are an essential part of creating a more unified labour market across the NP. Facilitating  knowledge spillovers requires arrangements that encourage and are conducive to the like minded entrepreneurs, researchers, politicians and business people  meeting and sharing ideas.  So far as subsidiaries are concerned the time profile of development could be an early dependence on externally owned firms setting up subsidiaries to exploit local advantages followed by a generation of  more locally owned, more autonomous subsidiaries as part of the development of more specialised supply trains in support of more innovative firms.  Manchester in particular, can look back on a long tradition of  working in this way with its innovations in railway transport in the early 19th century but also in its organisation of support for the interaction of science and business in the 1840s \citep{kargon1977science}.  The organisations that led that process remain in existence today in the form of the Literary and Philosophical Societies of Liverpool, Manchester,  Leeds, Sheffield, Newcastle and Hull together with many of the associated specialist groups. 
    
    According to  \cite{NESTA2016} there is still a need for boosting links between the city and entrepreneurs within the NP economy. NESTA recommends expansion of international transport linkages to strengthen global business connections. However, an analysis of the evolution of these business connections since the launch of the NPS in 2010 was not conducted in NESTA\textquotesingle s report. This paper seeks to fill this gap in the literature through a temporal assessment of the NPS within the system of cities framework, where preferential interactions between certain cities can give rise to the emergence of national sub-systems \cite{Pumain1997} -- in this  case, the economy of the North of the England. 
    
    The spring 2016 report on the NPS by HM Government \citep{HMGovernment2016} focuses mainly on transport issues and future road investments connecting NPS cities, but does not highlight the state of business connections between these cities. In the same perspective The Northern Powerhouse Independent Economic Review \citep{TheNorthernPowerhouseIndependentEconomicReview2016} identifies poor transport connectivity in the North as a major problem. Existing economic sector strengths are presented : advanced  engineering,  life  sciences  and  pharmaceuticals, chemicals,  energy  and  environment,  and  financial  and  professional  services. Agri-tech/food is a sector with notable growth potential in the North \citep{TheNorthernPowerhouseIndependentEconomicReview2016}. Yet no economic analysis of inter-urban interactions is conducted. The Greater Manchester Combined Authority (GMCA) industrial strategy released in June 2019 \cite{HMGovernment2019} also promotes global connectivity, supports innovative and international enterprise and foreign direct investments and sustains further internationalisation of the Greater Manchester business base. Four specific sectors are to be targeted in the future: advanced manufacturing (by creating the \say{Graphene City} at the University of Manchester and the \say{Advanced Materials City} in the North East of Manchester), health innovation, the already existing digital/media cluster and clean growth. 
    
    In this paper we wish to assess economic interactions within the NPS cities and the evolution of their economic sector capabilities within inter-urban ownership networks in the North of England. Literature on ownership networks is very scarce. Because of the lack of precise geo-located data, research is generally conducted either on aggregated data at regional or national level, on panel data for several years \citep{DemzetzLehn1985} or for one specific sector only \citep{FranksMayer1997,Bondetal2006, Wattsetal2007}. Studies on ownership links of firms at city level have been conducted for all types of sectors for different geographical areas \citep{Gillespie2017Finance, Martinusetal2019,Rozenblat2010,Rozenblatetal2017,RozenblatPumain2007,Sleszynski2015,Zdanowskaetal2020,2020Zdanowska} with the objective of modeling growth of urban firm networks \citep{raimbault2020modeling} . Ownership and industrial linkages have already been studied in Northern England, but for regions only \citep{Marshall1979}.

	We wish to identify polarisation and peripheralisation \citep{kuhn2015peripheral,benedek2015economic} effects within ownership links at firm level in Northern England. Polarisation is \textquotesingle the attraction exerted by a place on a more or less extended and heterogeneous one that is in a situation of dependency with respect to this centre \textquotesingle \citep{Elissalde2004}. This polarisation of the economy is the main driver of the peripheralisation effect \textquotesingle describing the production of peripheries through social relations and their spatial implications \textquotesingle \citep{kuhn2015peripheral}. This effect is particularly relevant in the case of the United Kingdom, given its centralised structure \citep{Arcauteetal2016}.
	
	\section*{Hypotheses}
	
	The hypotheses that we seek to test within this paper are:
	\begin{itemize}
	  \item {Hypothesis 1: The stated central and local policies for the NP have given rise to an economic sub-system in the North of the UK observable at the level of inter-urban interactions between firms;}
  
 	  \item{Hypothesis 2: The economic structure of the Northern cities taken together, offers a counterweight to the dominance of London and contributes to a rebalancing of the centralised economy of the UK;}
 	  
 	  \item{Hypothesis 3: The post 2010 industrial structure of NPS cities shows a shift away from a traditional manufacturing base towards more advanced technologies and diversification.}
	    
	\end{itemize}
	
	\newpage

	\section*{Materials and Methods}
	
    In order to test these hypotheses, we use:
    
    \begin{itemize}
    
    \item {The population and boundaries of UK cities in 2018 defined as 138 Functional Urban Areas with a common and harmonized definition of cities from the Global Human Settlement (GHS) database produced by the European Commission \citep{Schiavinaetal2019}.} 
    
    \item {The ownership links of capital between firms in 2010, 2013 and 2016 at city level from the ORBIS dataset \footnote{  The authors thank professor C{\'e}line Rozenblat -- Institut de g{\'e}ographie et durabilit{\'e}, Universit{\'e} de Lausanne (IGD - UNIL) -- for providing access to the data.}
    \cite{ORBIS2010,ORBIS2013,ORBIS2016} and from AMADEUS for 2018\footnote{Accessed via University College London (UCL) Library Services. AMADEUS is equivalent to ORBIS as it comes from the same BVD source, AMADEUS being the extraction for European firms and ORBIS covering the worldwide scale} \cite{AMADEUS2018}. Produced by the Bureau Van Dijk (BVD), both products indicate the exact geolocation, turnover and activity codes for owners and owned companies in all sectors. Data is available for UK companies and for the ones they own in the UK or overseas, as well as for UK companies owned by companies outside the UK. For 2010, the ORBIS dataset is composed of 203,403 links and 136,664 establishments, for 2013 of 186,321 links and 126 140 establishments and for 2016 of 232,620 links and 153,617 establishments. In AMADEUS 266,243 links and 289,808 establishments are available for 2018.}
    
    \end{itemize}
    
    The value of exploring Bureau Van Dijk\textquotesingle s products, is that its variables permit the reconstruction of ownership links between cities where companies are localised. For the purpose of analysing specifically the city of Manchester within the NPS, a decomposition of the ownership links was carried out and led to the identification of capital control chains at several levels, according to the following scheme: a firm (level L1) either in Manchester, London, in another UK city, in a city of the European Union or in a city outside Europe controls the capital of a firm in Manchester (level L2). The latter itself owns the capital of another firm (level L3) in Manchester, London, other cities of UK, in the European Union or outside Europe [see 3. in Figure 1].
    
    		\begin{figure}[H]
 \centering
 \includegraphics[width=1.0\textwidth]{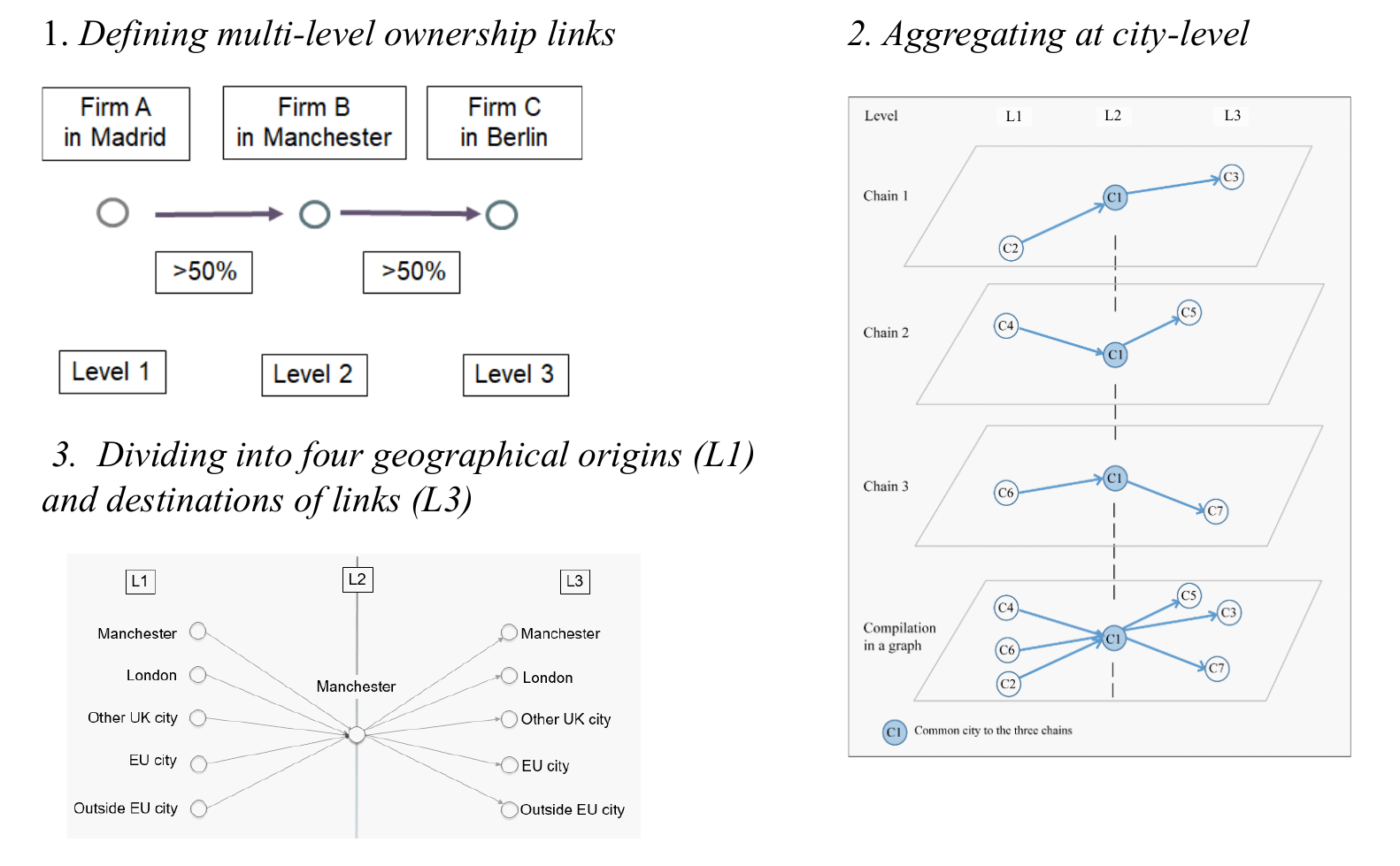}
  \caption{Defining the economic interurban ownership network of firms}\label{fig:1}
\end{figure}

    The three-level subnetwork of ownership networks in UK contains 2,312 firms and 1,562 ownership linkages. We have then calculated an index of intensity of the revenues generated by these ownership links at city level which we call \textquotesingle ownership links revenues\textquotesingle. The latter is proportional to the share of capital owned by the foreign firm and the turnover generated by the owned firm, expressed at the level of the city where the owned firm is located \citep{Sleszynski2015}. An aggregation of the chains, passing through the same city at level L2, was carried out according to all ownership links revenues generated in the given city [see 2 in Figure 1]. 

    An aggregation into 21 first level NACE \footnote{Statistical Classification of Economic Activities in the European Community (in French  "Nomenclature Statistique des activit\'es \'Economiques dans la Communaut\'e Européenne"} classification of activities has been applied for all analysis\footnote{(A) Agriculture, Forestry and Fishing, (B) Mining and Quarrying, (C) Manufacturing, (D) Electricity, Gas, Steam and Air Conditioning Supply, Water Supply; (E) Sewerage, Waste Management and Remediation Activities, (F) Construction, (G) Wholesale and Retail Trade; (H) Repair of Motor Vehicles and Motorcycles, Transportation and Storage, (I) Accommodation and Food Service Activities, (J) Information and Communication, (K) Financial and Insurance Activities, (L) Real Estate Activities, (M) Professional, Scientific and Technical Activities, (N) Administrative and Support Service Activities, (O) Public Administration and Defence; Compulsory Social Security, (P) Education (Q) Human Health and Social Work Activites (R) Arts, Entertainment and Recreation, (S) Other Service Activities, Activities of Households as Employers; (T) Undifferentiated Goods and Services Producing Activities of Households for Own Use (U) Activities of extraterritorial organisations and bodies.}\cite{EUROSTAT2008}. Correspondence analysis is a  multivariate statistical technique which summarises a set of categorical data in two-dimensional graphical form \cite{Hirschfeld1935,Benzecri1973,Greenacre1983}. It was applied in social science for the first time by  \cite{Bourdieu1984} and more recently in urban science for the analysis of trajectories of cities \citep{PaulusPumain2000}. In this paper correspondence analysis was run to characterise the economic ownership specialisation of cities based on the NACE 21 first level classification. The matrices used are described in the relevant results subsections. 

	\section*{Data / Empirics}
	
	\subsection*{Geographical scope of ownership}
	
	Understanding the geographical scope of ownership of firms is of economy-wide importance, as firms\textquotesingle  market value is equivalent to the global GDP \citep{DeLaCruzetal2019}. The geographical scope of ownership in the United Kingdom is mainly nationally driven and dominated by the capital city. This is a classical structure seen in many other countries \citep{LaPortaetal1999}. In 2016, 75 percent of firm owners in Manchester came from the UK and in 45 percent of cases from Manchester itself [Figure 2].
	
    		\begin{figure}[H]
 \centering
 \includegraphics[width=1.0\textwidth]{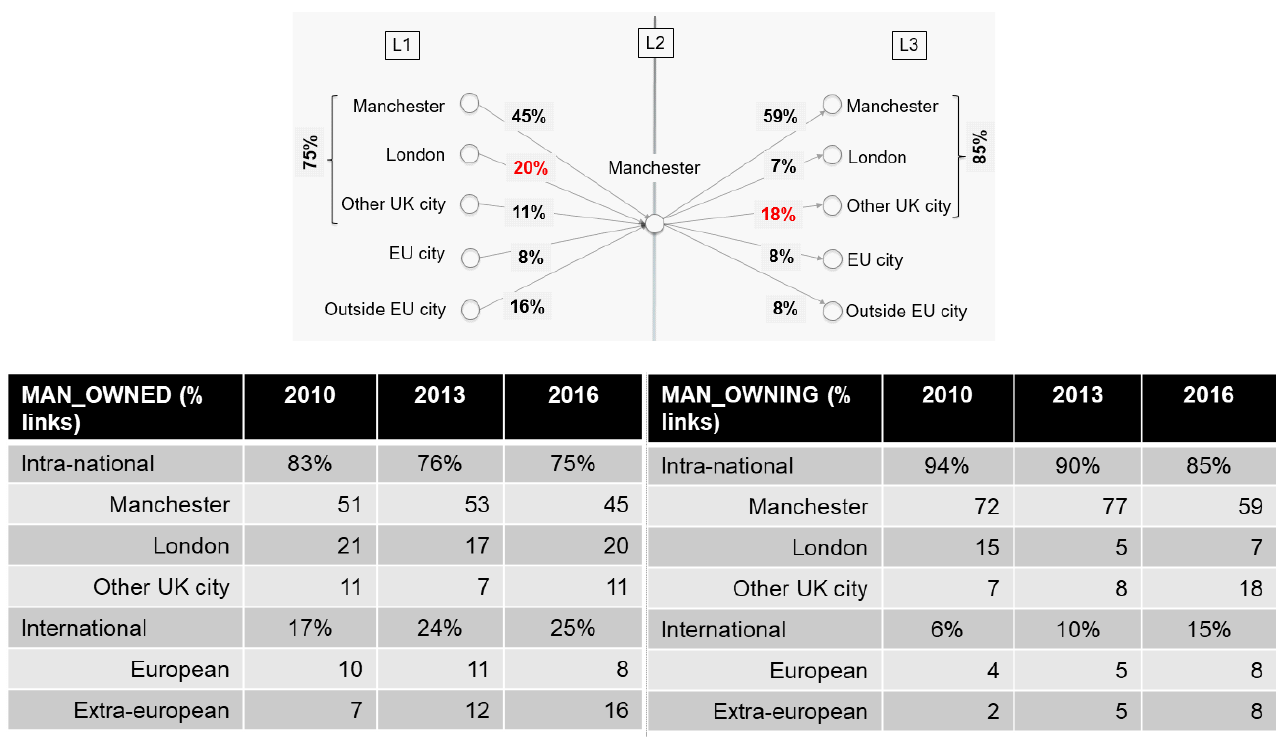}
  \caption{Multi-level ownership structure in 2016 and evolution since 2010; Man-owned -- Firms in Manchester owned by other firms; Man-owning -- Firms in Manchester owning other firms}\label{fig:1}
\end{figure}
	
	The Manchester firms\textquotesingle ownership structure for 2016 is oriented in 85 percent of cases towards the United Kingdom, although this has slightly diminished since 2010 (94 percent) and has become more internationally driven. In addition, in 2016, 18 percent of links coming from Manchester are oriented towards other UK cities. This orientation has more than doubled since 2010 in terms of number of links and at the same time the weight of London has diminished too. This relevant result will be investigated in more detail below  concentrating on the interactions between cities covered by the NPS. We aim at testing if the NP region defined by the government presents the potential characteristics of a regional subsystem in the United Kingdom (hypothesis 1) and whether it offers a counterweight to the dominance of London (hypothesis 2).
	
	\subsection*{Evolution of NPS inter-urban interactions}
	
	This section analyses the evolution of NP interactions within the UK. Firstly we consider inter-urban interactions within the NP. The total number of links between cities covered by the NPS –- here Manchester, Liverpool, Newcastle and Leeds \footnote{Other cities such as Hull and Sheffield covered by the NPS are not included due to the lack of ownership data} -– has been relatively stable through time: 376 in 2010, 303 in 2013 and 400 in 2016. However, when considering ownership links revenues generated from inter-urban ownership they have diminished through time. They initially rose from 18 billion euros ownership links revenues in 2010 to 24 billion in 2013 and then dropped to 9.5 billion in 2016 [Figure 3]
	\newline
	\footnote{\begin{itemize}
	   \item {Top 5 interactions of firms in Liverpool owning firms in Leeds in 2016: Energy Acquisitions UK Limited owns 100 percent of Dyneley Holdings Limited (force 1291788), Kingspan Investments Limited, owns 100 percent of Kingspan Access Floors Holdings Limited (force 100655), Heweston Holdings Limited owns 100 percent of Kingspan Access Floors Limited (force 58164), Thermal Ceramics UK Limited owns 34.96 of Jemmtec Limited (force 5978.859), Phoenix Medical Supplies Limited owns 21.1 of RX Systems Limited (force 4987.196)}
	  
	  \item{Top 5 interactions of firms in Newcastle owning firms in Leeds in 2016: DB (UK) Investments Limited owns DB Cargo (UK) Holdings Limited (force 768508), Arriva UK Trains Limited owns 100 percent of Arriva Trains Northern Limited (force 399709), Brompley Property Investments Limited owns 100 percent of BPT Limited (force 157283.7), Crown Packaging UK PLC owns 100 percent of CarnauxMetalBox Engineering Limited (force 120023), Nobia Holdings UK Limited owns 100 of Gower Group Limited (force 75130).}
	  
	  \item{Top 5 interactions of firms in Manchester owning firms in Newcastle in 2016: Rolls Royce Industries Limited owns 100 percent of Rolls Royce Power Engineering PLC (force 1044233), GE UK GROUP owns 41.22 percent of PII Group Limited (force 42771.52, nace 9411), LF Europe Limited owns 100 percent of Just Jamie and Paul Rich Limited (force 35429), Lornamead Group Limited owns 100 percent of Natural White (UK) Limited (force 18903), IGE USA Group owns PII Group Limited (force 9110.479), Sterling Fluid Systems (UK GROUP) limited owns Darwins Holdings Limited, force 100.}
	  
	  \end{itemize}}.

 	\begin{figure}[H]
    \centering
   \includegraphics[width=0.7\textwidth]{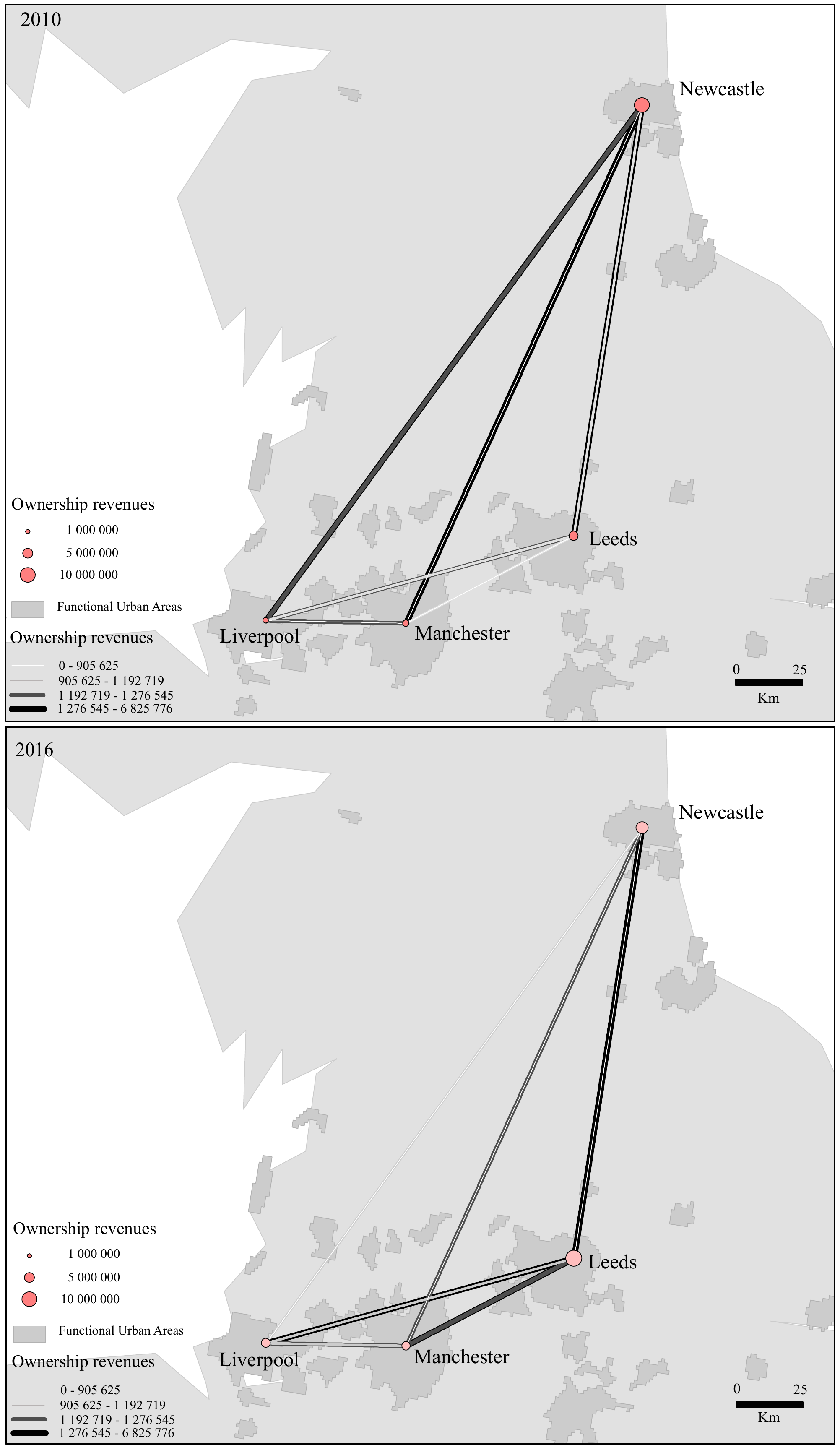}
   \caption{Ownership links between Manchester, Liverpool, Newcastle and Leeds in 2010 and 2016}\label{fig:2}
  \end{figure}

    		\begin{figure}[H]
 \centering
 \includegraphics[width=0.7\textwidth]{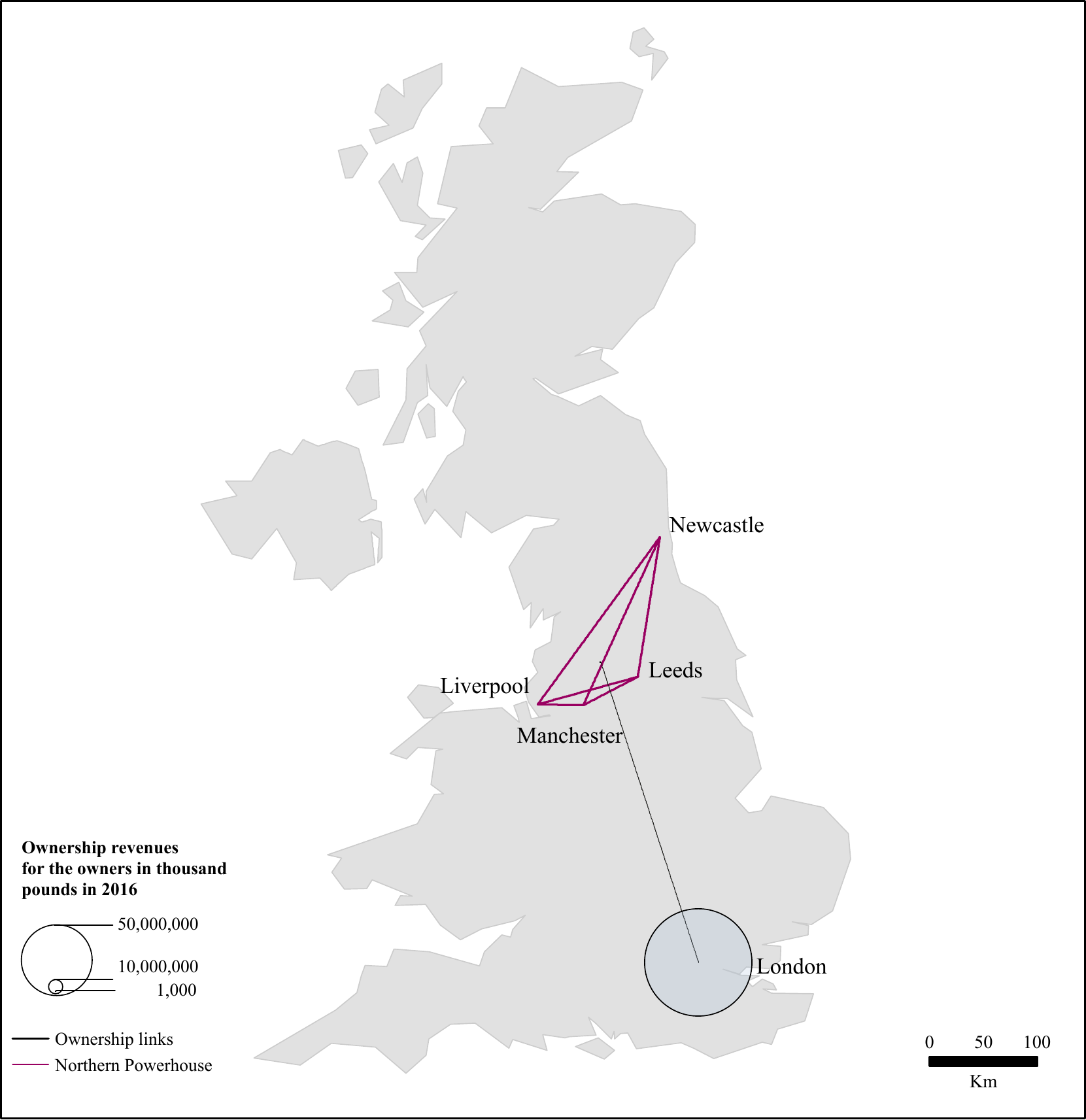}
  \caption{Ownership revenues generated by London-NP and NP-London ownership links in 2016}\label{fig:4}
\end{figure}
    
        \begin{figure}[H]

 \centering
 \includegraphics[width=0.6\textwidth]{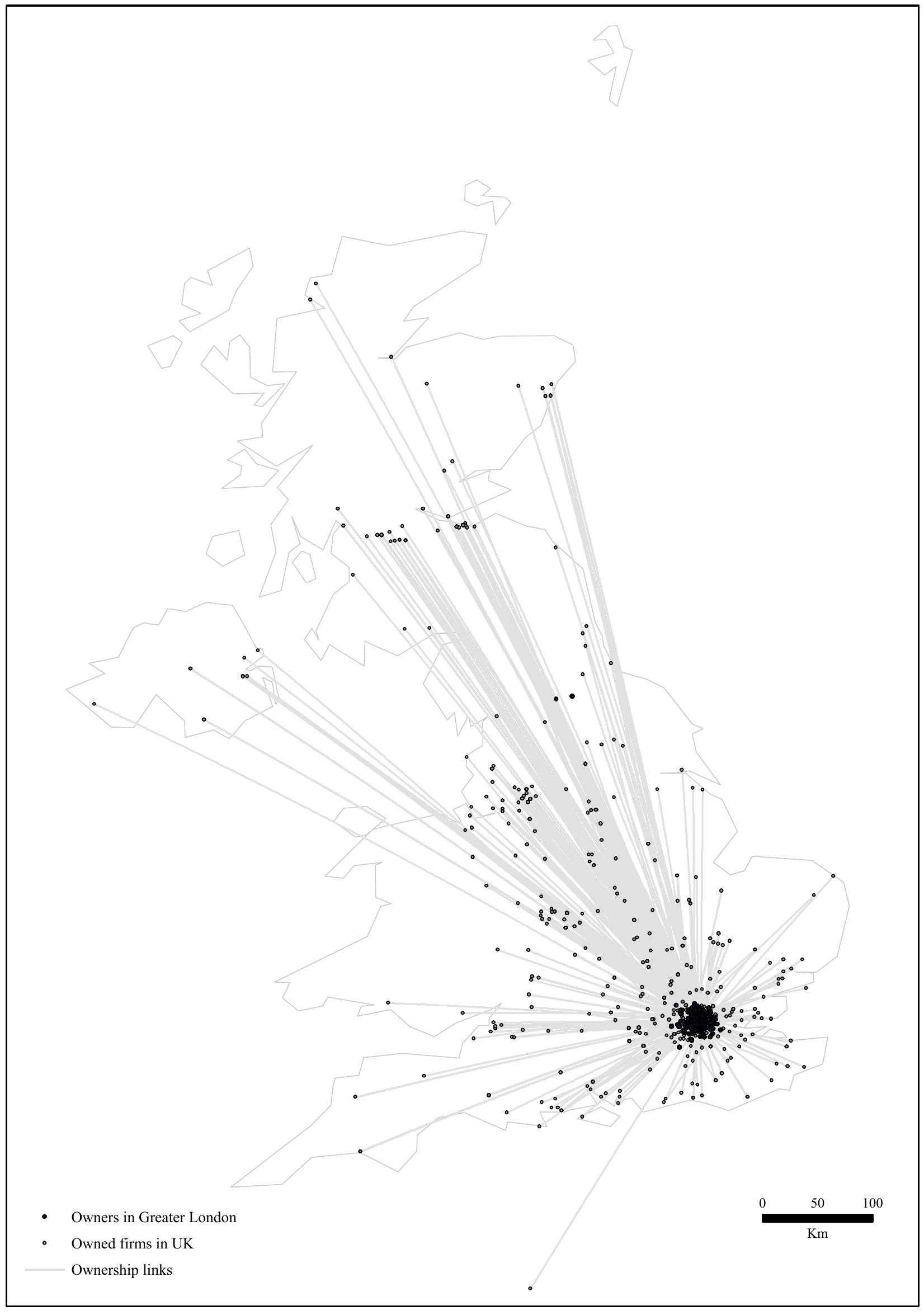}
  \caption{Ownership of firms in the United Kingdom by firms based in London in 2018}\label{fig:5}
\end{figure}

    The ownership strength between the NP cities has diminished probably due to an intensification of links oriented towards London. In terms of the initial hypotheses (1 and 2), it would seem that there has been little impact of the policy since 2010 in terms of creating an economy of the North in order to re-balance the UK economy and facilitate decentralisation. 
	
    When observing the ownership links coming from cities covered by the NPS and oriented towards other UK cities in 2018 [Figure 4], the NP economy clearly does not offer a counterbalance to the weight of firms in London. In fact the later generate more than 50 billions pounds in revenues coming from ownership in the NP, whereas firms in NP cities generate only 1 million pounds of revenues from ownership of firms in London. Firms in London dominate the whole UK territory in terms of ownership in 2018 [Figure 5]. To conclude this section, we reject hypotheses 1 and 2 and confirm our statement regarding the NP economy as being still far away from its initial objectives. In fact, the 2019 Centre for Cities report claims that NPS implementation is a job not even half done yet \cite{Centreforcities2019}. These results confirm that polarisation and peripheralisation effects are still ongoing in the United Kingdom. In the last section we will test our third hypothesis regarding the evolution of the post 2010 industrial base of NP cities. 
    
	\subsection*{Economic specialisation of cities within ownership networks}
	
	In this section we wish to examine the evolution of the economic specialisation of NP\textquotesingle s cities involved in ownership links between firms since 2010. At the same time this section will also verify the impact of the industrial strategy in terms of specialisation in advanced manufacturing, health, digital/media, clean growth, which are highlighted as sectors to be focused on in the future by the Combined Authorities of the North and by the NP.
    
    In order to determine the main economic specialisation of owned firms in Manchester, Liverpool, Newcastle and Leeds, a correspondence analysis has been conducted using the \textit{ade4} and \textit{explor} packages in R \citep{DrayDufour2007}. The analysis has been applied to a matrix of cities, in rows, and the economic sectors of the companies owned in these cities in columns, all expressed in total ownership links revenues by city and given sector [Figure 6].
    
		\begin{figure}[H]
 \centering
 \includegraphics[width=0.8\textwidth]{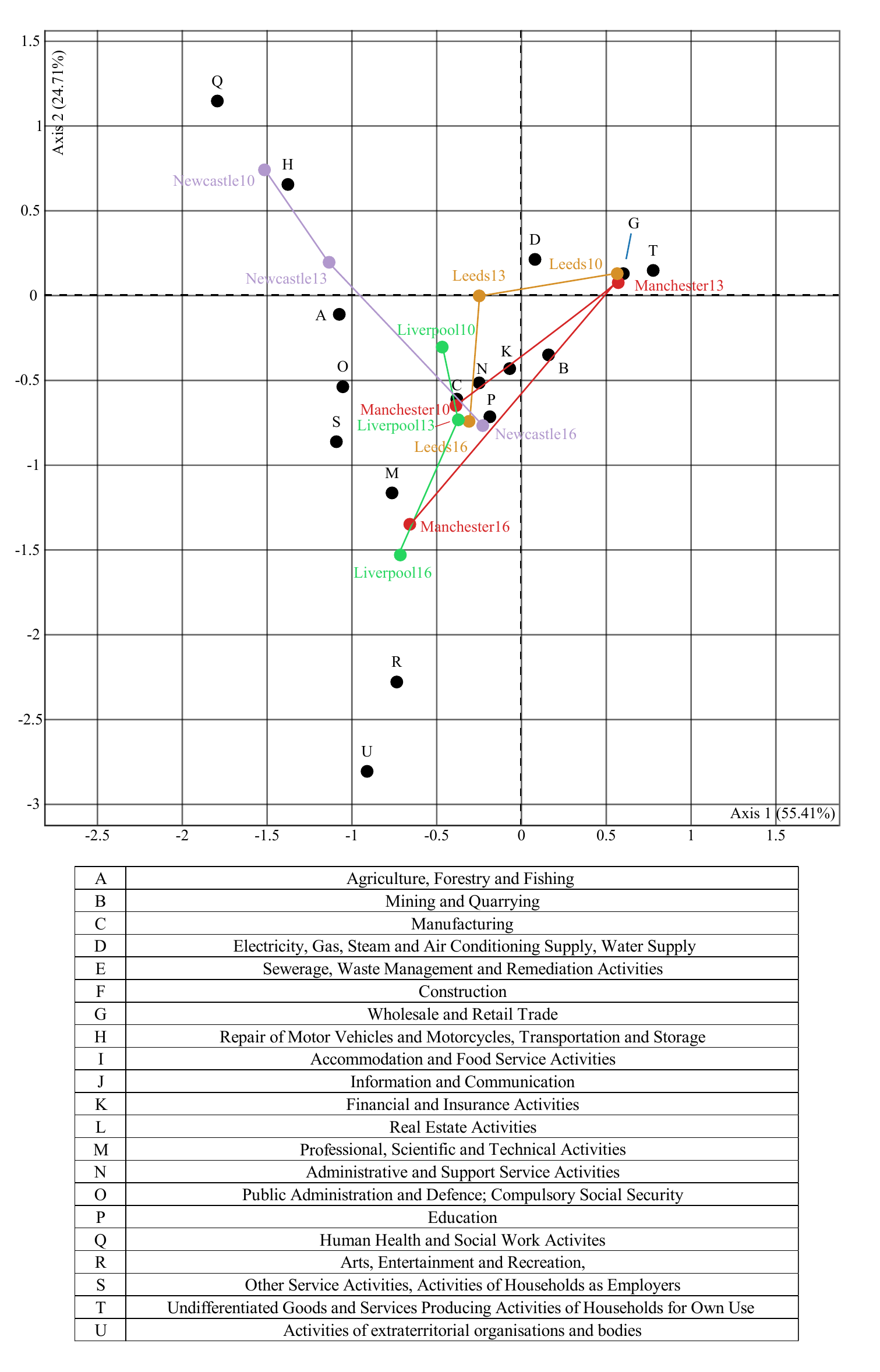}
  \caption{Trajectories of the NP cities in terms of their ownership specialisation in 2010, 2013, 2016}\label{fig:6}
\end{figure}

    The evolution of the trajectories of the four cities of the NP indicate a general upgrade in stages, of the domestic value chain \citep{Cadestinetal2019}, as the links coming from other firms have become more specialised in advanced manufacturing, research and development and health. This justifies the priorities of the NPS for the future and corresponds to the economic sector capabilities of Northern cities as defined by the NPS and the GMCA\textquotesingle s industrial strategy.
    
    The correspondence analysis shows that firms owned in Manchester have switched from specialisation in manufacturing (C), to wholesale and trade (G) and finally to professional scientific and technical activities (M) in 2016. This confirms the GMCA\textquotesingle s industrial strategy as investments in scientific activities are related to the Manchester University strategy for the production of graphene as well as the creation of a Graphene City in the North \citep{HMGovernment2019}. In fact Manchester now presents a strong advanced manufacturing base in materials and textiles, chemicals, food and drink and is the site of Trafford Park, one of Europe’s largest industrial parks \citep{HMGovernment2019}.
    
    This switch in the global value chain is particularly relevant in the case of Liverpool as firms owning capital have converted from Agriculture, Forestry and Mining (A) in 2010 to Professional scientific and technical activities (M) in 2013. Ownership in Newcastle switched from repair of motor vehicles and motorcycles (H) to agriculture, forestry and mining (A) to education (P). Leeds went from a specialisation in wholesale and trade (G) to electricity, gas and steaming (D) to manufacturing (M) and education (P). This is a relevant observation as it shows that owners of firms invest in scientific activities and health as a priority and these sectors are the ones generating the most revenues. These results weaken the findings in the previous section relating to hypotheses 1 and 3 and confirm that there has been a shift away from a traditional manufacturing base towards a more diversified specialisation of cities and this might in part, be due to the impact of the NPS since 2010.
    
    However when coming back to hypothesis 2 the opposition between the Northern cities and London is still very important in 2018  [Figure 7]. On one side London concentrates firms owned in all kinds of sectors abut mainly Finance and Insurance activities (K), whereas on the opposite side of the factorial axis Leeds, Newcastle and Liverpool show specialisation in retail and electricity, gas, steam and air conditioning supply. There is still an enormous gap to fill and initiatives needed in order to create a stronger economic subsystem in the North capable of counterbalancing the weight of London and to meet the foresight for cities objectives in terms of diversification of the economic base in order to reduce polarisation and peripheralisation of the North of England. 

		\begin{figure}[H]
 \centering
 \includegraphics[width=0.9\textwidth]{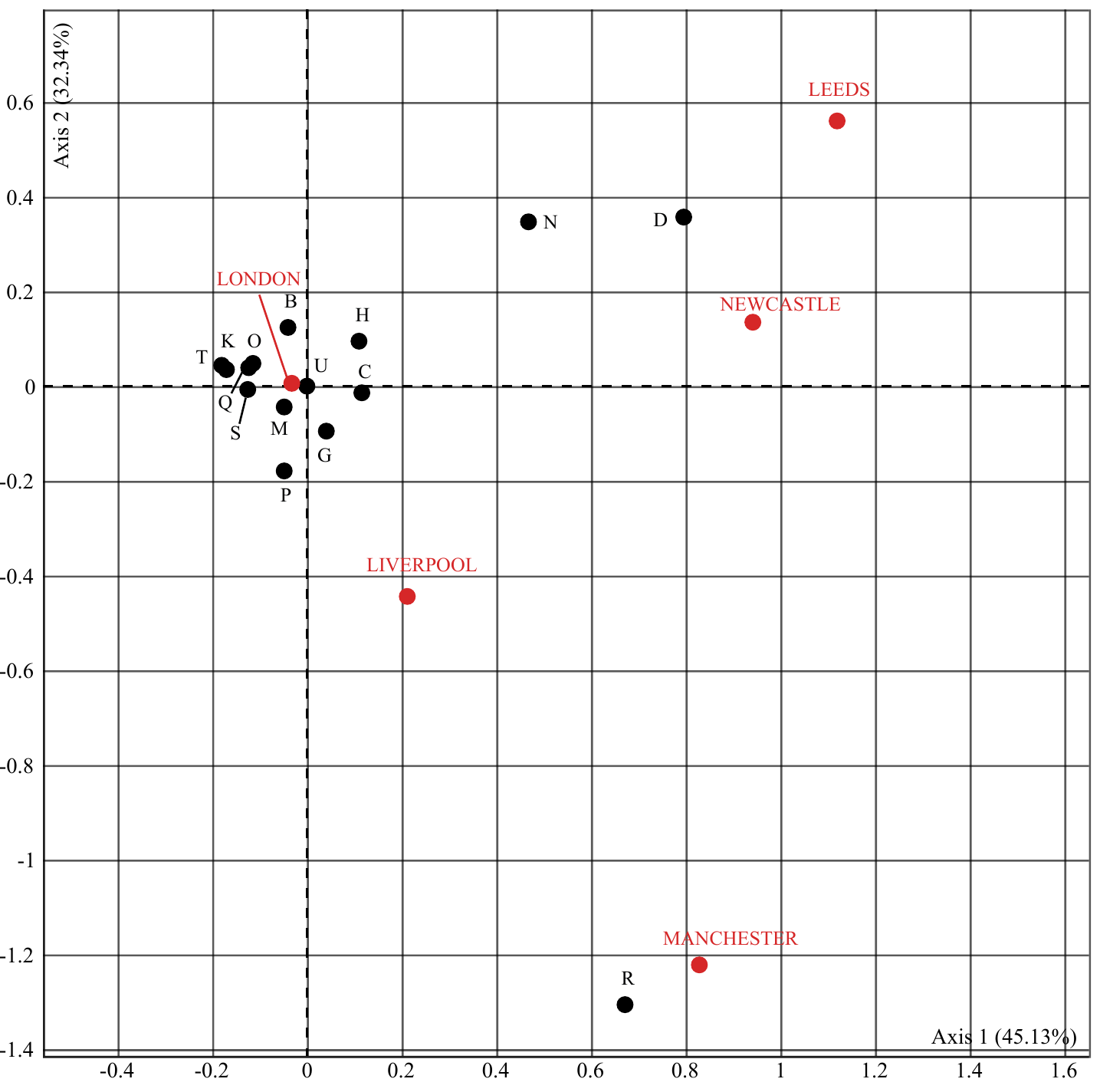}
  \caption{NP cities and London in terms of their ownership specialisation in 2018}\label{fig:4}
\end{figure}

    \section*{Discussion}

The analysis firstly shows that the geographical scope of ownership of firms located in Manchester is mainly oriented towards firms in the city itself or in London. Local links towards other UK cities outside London represent only 18 percent of total links in 2016. 
Secondly the analysis of the evolution of these local links between Manchester, Liverpool, Leeds and Newcastle has revealed a decrease of revenues generated by inter-urban ownership links and a relatively stable number of interactions since 2010. The latter were polarised on Liverpool, Manchester and Leeds leading to a peripheralisation of Newcastle in 2016. 
Thirdly, findings show that Liverpool and Manchester have switched from a traditional manufacturing base towards a specialisation in professional and scientific activities within the ownership networks in this same period. Despite the modernisation of the Manchester and Liverpool industrial structure, it would seem that the NPS has had only a minor impact on economic interaction between Northern cities since 2010. 

This paper shows that the NPS has not yet created an integrated sub-system between Northern cities of the United Kingdom -- at least from the economic perspective of ownership links between firms -- and is still characterised by peripheralisation and polarisation patterns when compared to London. The NP is still far from contributing to  decentralisation and foresight for cities defined by the government. The evolution of the economic specialisation of Liverpool and Manchester in the global value chain of production reveals that this process of creating an economy of the North capable of withstanding the effect of London is still ongoing. These findings are relevant for policy-makers regarding the cities\textquotesingle foresight government objectives for the next fifty years. An initiative such as the NP needs to be be continued and reinforced for the future of cities to create a more resilient and balanced national system of cities. A future extension of this research could include transportation and commuting interactions between cities of the NP, as these may well exhibit patterns of interaction that differ from those of the economic networks. Furthermore adding the city of Hull in the NPS analysis would bring a different perspective, as Hull presents a developing economic structure in the North of England. This research suggests that one may think differently about the NP from a network perspective with a focus on inter-urban interactions rather than focusing on within  city interactions. A further extension of this research would be its application to  case studies in other countries\textquotesingle city-systems, where decentralisation is a major issue.

\section*{Acknowledgments}
    
The authors acknowledge support from the Urban Dynamics Lab under the EPSRC Grant No.  EP/M023583/1 

\bibliographystyle{apalike.bst}

\end{document}